\documentclass[twocolumn,showpacs,preprintnumbers,amsmath,amssymb]{revtex4}

\usepackage{graphics}
\usepackage{epsfig}

\usepackage{graphicx}
\usepackage{dcolumn}
\usepackage{bm}

\begin{document}

\title{Massless surface wave}

\author{I.\;Todoshchenko}
\affiliation{Low Temperature Laboratory,
Deptartment of Applied Physics, Aalto University University, 00076~AALTO, Finland}
\email{todo@boojum.hut.fi}
\date{\today}

\begin{abstract}
An interface between two media is topologically stable two-dimensional object where 3D-symmetry breaks
which allows for existence of many exotic excitations. A direct way to explore surface excitations is
to investigate their interaction with the surface waves, such as very well known capillary-gravity 
waves and crystallization waves.
Helium remains liquid down to absolute zero where bulk excitations are frozen out and do not mask the
interaction of the waves with the surface states. 
Here we show the possibility of the new, massless wave which can propagate along the surface between two 
different superfluids phases of $^3$He. The displacement of the surface in this wave occurs due to
the transition of helium atoms from one phase to another, so that there is no 
flow of particles as densities of phases are equal. We calculate the dispersion of the wave in which
the inertia is provided by spin supercurrents, and the restoring force is magnetic field gradient.
We calculate the dissipation of the wave and show the preferable conditions to observe it.
\end{abstract}


\maketitle

\begin{figure}[t]
\includegraphics[width=1.0\linewidth]{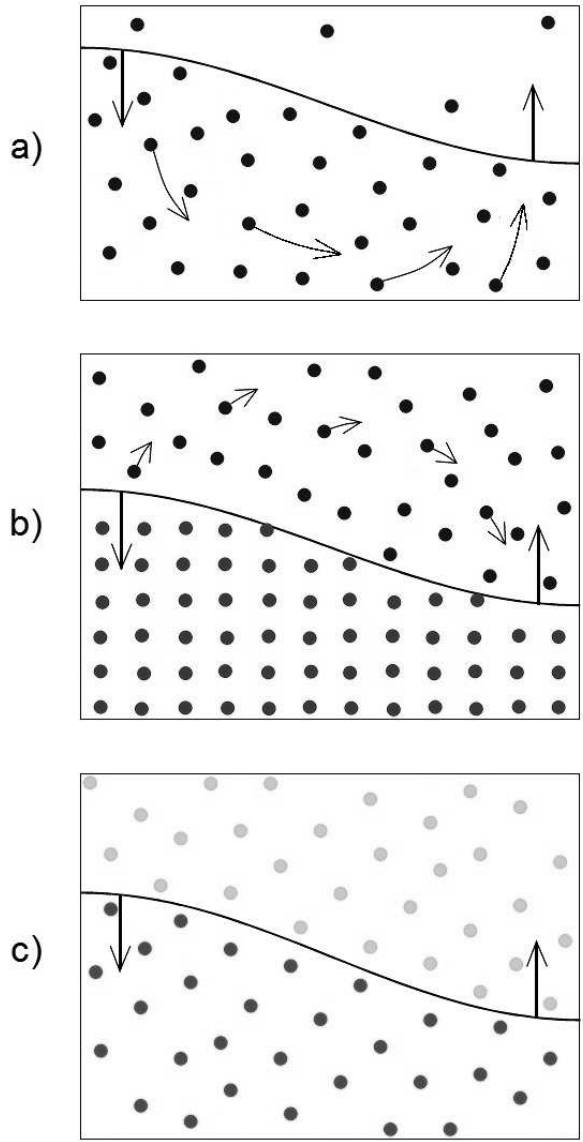}
\caption{\label{fig:3waves} Three types of surface waves. a) Usual capillary-gravity wave: oscillations
of the surface are accompanied with the motion of the liquid in the bulk. b) Crystallization wave:
oscillations of the surface are accompanied with the much slower motion of the liquid because density of
liquid and solid are similar. c) Phase wave on the surface between two superfluid phases of $^3$He:
no mass transport is needed because densities of both superfluid phases are equal.
}
\end{figure}

Watching the waves on a surface of ocean is probably one of the oldest physical observation. 
Similar waves can propagate along the surface between any two immiscible fluids,
and particularly between liquid and its vapour, see Fig.\;\ref{fig:3waves}a. The inertia of these waves is 
due to the motion of the liquid while the restoring force is due to gravity (long waves) and surface tension
(short waves). Unlike the most of known waves, dispersion relation of surface wave is strongly nonlinear,

\begin{equation}
\label{eq:capillary}
\rho \omega^2=\rho g q + \alpha q^3,
\end{equation} 

\noindent
so that there is no phonon-like spectra even at $q\rightarrow 0$. Here $\rho$ is the density of the liquid,
$\omega$ is the angular frequency, $g$ is the gravitational acceleration, $\alpha$ is the surface tension,
and $q$ is the wavevector. Capillary-gravity waves have been observed on surface of 
$^4$He and $^3$He and have been utilized for high-precesion measurements on the surface tension. In 
this way critical exponents near critical liquid-gas point have been measured \cite{Iino86}.

For helium liquid-solid interface there opens a possibility for quite exotic wave, a
crystallization wave. At low enough temperatures, where liquid phase is superfluid and thus
provides extremely fast mass and heat transport, the interface between solid and liquid becomes
mobile enough to support wave of crystallization and melting. The dispersion relation is
similar to that of waves on free liquid surface, and the only difference is that for the same amplitude
of the wave, liquid phase carries smaller mass flux (see Fig.\;\ref{fig:3waves}b)  which is proportional 
to density difference $\Delta\rho$ between the solid and the liquid,

\begin{equation}
\label{eq:crystallization}
\frac{\Delta\rho^2}{\rho} \omega^2=\Delta\rho g q + \alpha q^3.
\end{equation}

Crystallization waves have been predicted by Andreev and Parshin in 1978 \cite{AP} and discovered 
by Keshishev {\it et al}.~two years later in $^4$He below 0.5\;K \cite{Keshishev80}. By measuring 
crystallization waves at surfaces of different orientations the singularity of the surface tension at 
the basal c-facet orientation has been observed \cite{Rolley95}.

In this Letter we show the possibility of the new kind of surface wave which does not have any associated mass
flux. Imagine an interface between two immiscible liquids which both consist of the same 
atoms or molecules. Such interface will support, in addition to usual capillary-gravity wave described 
above, another type of wave which is associated with the transition of particles near the surface from one 
phase to another one. Basically, such wave is an analog of the crystallization wave as one of 
the phases locally ``grows'' into another phase. The hydrodynamical motion in both liquids 
is thus only due to the difference in mass densities of the phases.

Generally, it is possible that both phases have the same or nearly the same mass densities. Then such a wave 
will not have any hydrodynamical inertia and its frequency will be infinitely high at any finite wavevector
$q$ unless some other kind of inertia is involved. The simplest example of the discussed system
is liquid in contact with its vapour close to the critical liquid-vapour point where densities of the 
both phases equalize. Moreover, at the critical point surface tension also vanishes, and thus there is 
no restoring force unless some other force stabilizing the surface is introduced.

Although it seems impossible to satisfy all the conditions for the discussed massless wave,
nevertheless, there is an
example of such system, a surface between two superfluid phases of helium-3, stabilized with the magnetic 
field \cite{Bradley95,Eltsov02}. These two phases, being referred to as ``A'' and ``B'' phases, have
the same mass densities, and the described phase wave on the A-B interface does 
not require mass transport in the bulk phases, see Fig.\;\ref{fig:3waves}c. 

However, the A-phase has larger magnetic susceptibility and thus prefer large magnetic field. By applying 
vertical magnetic field gradient $\nabla{H}$ one can stabilize the A-phase in high field region on top of 
the B-phase. The field gradient is therefore an analog of gravity and plays a role of the restoring force 
of the wave.

As the magnetization of the A-phase is larger than the magnetization of the B-phase, the motion of the A-B 
interface is accompanied with the spin currents in both phases. As it was noted by Andreev, spin currents 
have a kinetic energy \cite{Andreev93}, and thus play the role of inertia. We therefore have a wave on the 
surface between two liquids in which there is no mechanical motion at all: $^3$He particles transit from 
one phase to another staying at rest, and both kinetic and potential energy of the wave are provided by 
the spin degree of freedom.

As it was shown by Leggett \cite{Leggett75}, the spin dynamics in the longitudinal direction can be described 
in the terms of the angle $\theta$ of rotation of the order parameter vector $\vec{d}$ 
around the field direction. The corresponding equations for the A-phase ($z>0$) and for the B-phase ($z<0$)
are \cite{Andreev93}:

\begin{equation}
\label{eq:dynamics}
\ddot{\theta_a}-c_a^2\frac{\partial^2\theta_a}{\partial z^2}+\Omega_a^2\theta_a=0,~~~~~~~~
\ddot{\theta_b}-c_b^2\frac{\partial^2\theta_b}{\partial z^2}+\Omega_b^2\theta_b=0,
\end{equation}

\noindent
Here $\Omega_a$, $\Omega_b$, $c_a$, and $c_b$ are longitudinal NMR frequencies and spin waves velocities of
A- and B- phases, correspondingly. We consider a slow wave, $\omega\ll\Omega_a,\Omega_b$, for which 
the solution of (\ref{eq:dynamics}) decaying at $z=\pm\infty$ is 

\begin{equation}
\label{eq:theta}
\theta_{a,b}(z,t)=\theta_{0a,0b}(t)\exp{\big(\mp\frac{\Omega_{a,b}}{c_{a,b}}z\big)}.
\end{equation}

\noindent
Following Andreev \cite{Andreev93}, we write the energy density of the spin waves as

\begin{equation}
\label{eq:density}
\varepsilon_{a,b}=\frac{\chi_{a,b}}{2\gamma^2}
[c_{a,b}^2(\frac{\partial\theta_{a,b}}{\partial z})^2 + \Omega_{a,b}^2\theta_{a,b}^2],
\end{equation}

\noindent
where $\chi_a$ and $\chi_b$ are magnetic susceptibilities of the phases, and $\gamma=2\mu/\hbar$ is the 
gyromagnetic ratio. After integration over $z$ we find the total kinetic energy of the wave,
\begin{equation}
\label{eq:kinetic}
E_{kin}=\frac{1}{2}\frac{\chi_ac_a\Omega_a}{\gamma^2}\theta_{0a}^2+
         \frac{1}{2}\frac{\chi_bc_b\Omega_b}{\gamma^2}\theta_{0b}^2.
\end{equation}

\noindent
The fluxes of the $z$ component of the spin at the boundary are given by

\begin{equation}
\label{eq:flux}
j_{a,b}=-\frac{\chi_{a,b}c^2_{a,b}}{\gamma^2}\frac{\partial\theta}{\partial z}=
\pm\frac{\chi_{a,b}c_{a,b}\Omega_{a,b}}{\gamma^2}\theta_{0a,0b}.
\end{equation}

\noindent
The fluxes are connected by the boundary condition

\begin{equation}
\label{eq:boundary}
j_a-j_b=\frac{H}{\gamma}(\chi_a-\chi_b)\dot{\zeta},
\end{equation}

\noindent
where $\zeta(x,t)=\zeta_0(t)\exp{(iqx)}$ is the displacement of the boundary from its equilibrium $z=0$ position.
Using (\ref{eq:boundary}) and minimizing kinetic energy (\ref{eq:kinetic}) we obtain
amplitudes of the spin currents,

\begin{equation}
\label{eq:amplitude}
\theta_{0a}=\theta_{0b}=
  \frac{\gamma H(\chi_a-\chi_b)}{\chi_ac_a\Omega_a+\chi_bc_b\Omega_b}\dot{\zeta}.
\end{equation}

\noindent
Substituting amplitudes (\ref{eq:amplitude}) to (\ref{eq:kinetic}) we find
 the kinetic energy of the wave, $E_{kin}=(1/4)M\dot{\zeta}^2$ with the effective 
``mass'' of spin currents

\begin{equation}
\label{eq:mass}
M=\frac{(\chi_a-\chi_b)^2H^2}{\chi_a c_a \Omega_a + \chi_b c_b \Omega_b}.
\end{equation}

\noindent
The potential energy of the wave is due to the field gradient and due to the small surface tension
$\alpha$, 
$E_{pot}=(1/4)[(\chi_a-\chi_b)H\nabla{H}+\alpha q^2]\zeta^2$, and the dispersion relation of the wave is

\begin{equation}
\label{eq:dispersion}
\omega^2=\frac{(\chi_a-\chi_b)H\nabla{H}+\alpha q^2}{M}.
\end{equation}

Note that in the limit of long waves frequency does not depend on the wave vector and is determined by only 
magnetic parameters of superfluid phases and by the geometry of the magnetic field. At low temperatures 
and low pressure the A-B interface is stabilized in the field of $H_{AB}\approx0.34$\;T \cite{Bartkowiak99}, 
and the surface tension is $\alpha\approx5$\;nJ$/$m$^2$ \cite{Haley02}. If we assume the field gradient of 
$\nabla{H}\sim10$\;T$/$m, then the characteristic wavelength at which the two terms 
in Eq.\;(\ref{eq:dispersion}) equalize is 

\begin{equation}
\label{eq:crossover}
\lambda_c\sim2\pi\sqrt{\frac{\alpha}{(\chi_a-\chi_b)H\nabla{H}}}\sim1\;{\rm mm}.
\end{equation}

\noindent
Longer waves are inherently magnetic, and their frequency $\omega_0\sim\sqrt{c\Omega\nabla{H}/H}$ does not 
depend on wavevector but can be tuned by the magnetic field gradient.

The wave can be exited by applying an oscillating magnetic field $h=h_0\exp{(i\omega t)}$ parallel to the 
constant stabilizing field $H$. This kind of experiment has been done in Lancaster by Bartkowiak {\it et al.}, 
who have measured the heat produced by the oscillating A-B interface in the ultra 
low temperature limit at zero pressure \cite{Bartkowiak00}. The oscillating field
$h$ causes the equilibrium vertical position of the interface to oscillate with the amplitude
$\delta\zeta=h/\nabla{H}$. The equation of motion of the interface can be written as

\begin{eqnarray}
\label{eq:motion}
M\ddot{\zeta}+\Gamma\dot{\zeta}=-H\nabla{H}(\chi_a-\chi_b)(\zeta-\delta\zeta\exp{(i\omega t)})= \nonumber \\
=-M\omega_0^2(\zeta-h/\nabla{H}\exp{(i\omega t)})~~~~~~~~~~~~
\end{eqnarray}

\noindent
where $\Gamma$ is the friction coefficient. The frequency dependence of the power dissipated in the wave
per unit area is given by 

\begin{eqnarray}
\label{eq:power}
P=\Gamma\overline{|\dot{\zeta}|^2}=\Gamma\omega^2\overline{|\zeta^2|}=~~~~~~~~~~~~~~~~~ \nonumber \\
=\frac{\Gamma}{2}\frac{\omega_0^4\omega^2}{(\omega_0^2-\omega^2)^2+(\omega\Gamma/M)^2}\frac{h^2}{\nabla{H}^2}.
\end{eqnarray}

The quality factor of the wave $Q=\omega_0M/\Gamma$ is proportional to the square root of the 
effective mass of the wave $M$ which is very small because of the smallness of nuclear 
susceptibilities, $\chi_a=3.8\cdot10^{-8}$, $\chi_b=1.2\cdot10^{-8}$ at zero bar \cite{Wheatley75}. 
The longitudinal NMR frequencies in the low temperature limit at high pressure have been measured to be 
$\Omega_a|_{hp}=2\pi\cdot100$\;kHz, $\Omega_b|_{hp}=2\pi\cdot250$\;kHz
\cite{OsheroffBrinkmann74,Webb74,Bozler74,Osheroff74}. 

As it was shown by Leggett, the longitudinal resonance
frequency is scaled as $\Omega\propto\sqrt{N(0)a}T_c$ where $N(0)$ is the
density of states at the Fermi surface, and $a$ is the ratio of the relative specific heat jump to its
BCS value (1.43) \cite{Leggett72}. Using $N(0)|_{hp}=1.26\cdot10^{51}$\;J$^{-1}$m$^{-3}$, 
$N(0)|_{0bar}=0.54\cdot10^{51}$\;J$^{-1}$m$^{-3}$ \cite{Wheatley75}, $a|_{hp}=1.4$, $a|_{0bar}=1$,
$T_c|_{hp}=2.49$\;mK, and $T_c|_{0bar}=0.93$\;mK \cite{Halperin76,Alvesalo80,Greywall86} we find
$\Omega_a|_{0bar}\simeq2\pi\cdot25$\;kHz, and $\Omega_b|_{0bar}\simeq2\pi\cdot65$\;kHz. 
For the estimation of the velocity $c$ of spin waves one can use the relation 
$c=l(\Omega/2\pi)$ where $l\approx10\;\mu$m is the dipole healing length. Finally, we find the mass of
the wave at high pressures and at zero bar, $M|_{0bar}\approx8\cdot10^{-9}$\;kg$/$m$^2$ and 
$M|_{hp}\approx2\cdot10^{-9}$\;kg$/$m$^2$. 

Another kind of the inertial mass $M^*$ of the interface originates from the time variation of the 
order parameter near the moving interface. This mass has been first considered by Yip and Leggett who have 
given a rough estimate $M^*\sim10^{-11}$\;kg$/$m$^2$ \cite{Leggett86,Leggett90}. Roughly, this mass 
is smaller than the magnetic mass by the factor $\xi_0/l$, where $\xi_0=18\,...\,88$\;nm is the coherence
length.

There also exists usual material mass of the wave because of the density difference 
$\Delta\rho_{AB}\sim10^{-6}$\,kg$/$m$^3$ between phases \cite{Leggett90}. The applied magnetic field 
increases the density difference because the A-phase has large susceptibility. The magnetostriction effect 
can be estimated as the additional effective pressure $\delta P=(1/2)\chi H^2$ which causes compression of 
the liquid by an amount $\delta\rho/\rho=\beta\delta P$; where $\beta=5.7\cdot10^{-8}$\;1$/$Pa is the 
compressibility \cite{Grilly}. The additional density difference due to magnetostriction is thus 
$\Delta\rho_{AB}(H)=(1/2)(\chi_A-\chi_B)H^2\beta\rho\sim10^{-8}$\;kg$/$m$^3$ which is negligible
compared to zero field difference $\Delta\rho_{AB}$. The contribution to the mass of the wave from
density difference of the phases is given by $m=\Delta\rho_{AB}^2/(\rho q)\sim10^{-17}$\;kg$/$m$^2$
which is nine orders of magnitude smaller than the spin current mass $M$.

With the estimated above mass $M\sim10^{-8}$\;kg$/$m$^2$ and with the field gradient of few Tesla per meter
we can estimate the resonant frequency of the wave $\omega$ to be of the order of $2\pi\cdot1$\,kHz.
This value is indeed much smaller than the longitudinal NMR frequency, and thus the condition of slow
wave used for calculations of the mass is valid. In the opposite case of fast wave (very short wavelength
and/or very strong field gradient), the magnetizations of the bulk phases will not have time to reach 
their equilibrium values, and both the solution (\ref{eq:theta}) and the boundary condition 
(\ref{eq:boundary}) fail. For this case more complicated analysis is needed.

The friction coefficient $\Gamma$ is determined by scattering of quasiparticles in the A-phase and decreases 
at ultra low temperatures as $T^4$ \cite{Leggett86,Kopnin87}. It has been measured at temperatures close 
to the A-B transition at high pressures in Los Alamos \cite{Buchanan86}, 
$\Gamma|_{0.75T_c,hp}=0.07\;{\rm kg/(m^2s)}$. This value is in very good agreement 
with theoretical estimation of Kopnin, $\Gamma\approx 7\pi^4 N(0)T^4/(30v_F\Delta_A^2)$ ($v_F$ is
Fermi velocity, $\Delta_A$ is the BSC energy gap) \cite{Kopnin87}. 
The low temperature value of the friction coefficient 
can be found from the Lancaster experiment \cite{Bartkowiak00}. According to the Eq.\;(\ref{eq:power}), in the 
case of strong damping $\Gamma/M\gg\omega_0$, the dissipation has a plateau at high frequencies, 

\begin{equation}
\label{eq:plato}
P_{pl}=\frac{1}{2\Gamma}\frac{\omega_0^4 M^2 h^2}{\nabla{H}^2}=
h^2 H^2 (\chi_a-\chi_b)^2\frac{1}{2\Gamma},
\end{equation}

\noindent
which depends only on the amplitude $h$ of the oscillating field and on the friction coefficient $\Gamma$.
Indeed, such plateau has been observed in Lancaster at frequencies larger than 10\;Hz with the level 
$P_{pl}=2\cdot10^{-7}$\;W$/$m$^2$ independent on the field gradient \cite{Bartkowiak00}. The amplitude
of the oscillating field used in Lancaster experiment was $h=0.64$\,mT which gives
$\Gamma|_{0.17T_c,0{\rm bar}}=3\cdot10^{-5}\;{\rm kg/(m^2s)}$. This is about three times larger
than the theoretical estimate.

Due to the smallness of the mass of the wave, the characteristic attenuation rate of the wave is very 
fast even at ultra low temperatures, 
$\Gamma|_{0.17T_c,0{\rm bar}}/M\approx4\cdot10^3$\;s$^{-1}$ which should be compared to
$\omega_0\approx1.4\cdot10^3$\;s$^{-1}$ for the strongest field gradient $\nabla{H}=2$\;T$/$m used
in Lancaster experiments. To observe resonance of the A-B wave one should increase the field gradient
by at least order of magnitude to shift the resonant frequency $\omega_0$ above the dissipation rate.

The excess by factor of three of the dissipation measured in Lancaster over the theoretical value
might be the indication of the contribution of the surface states which should 
dominate at zero temperature limit. The A-B interface between two fermionic superfluids  is 
probably the richest surface in nature and promises to support variety of surface states such as Majorana 
fermions and anyons (particles which are neither bosons nor fermions) \cite{Stern10}. The proposed magnetic 
massless surface wave could open access to these exotic surface states which may cause additional dissipation 
and contribute to the mass of the wave.

\begin{acknowledgments}
I am grateful to Vladimir Eltsov and Grigory Volovik for valuable discussions.
This work has been supported by the Academy of Finland (LTQ CoE grant no.~250280).
\end{acknowledgments}


\begin{thebibliography}{26}
\expandafter\ifx\csname natexlab\endcsname\relax\def\natexlab#1{#1}\fi
\expandafter\ifx\csname bibnamefont\endcsname\relax
  \def\bibnamefont#1{#1}\fi
\expandafter\ifx\csname bibfnamefont\endcsname\relax
  \def\bibfnamefont#1{#1}\fi
\expandafter\ifx\csname citenamefont\endcsname\relax
  \def\citenamefont#1{#1}\fi
\expandafter\ifx\csname url\endcsname\relax
  \def\url#1{\texttt{#1}}\fi
\expandafter\ifx\csname urlprefix\endcsname\relax\def\urlprefix{URL }\fi
\providecommand{\bibinfo}[2]{#2}
\providecommand{\eprint}[2][]{\url{#2}}

\bibitem[{\citenamefont{Iino et~al.}(1986)\citenamefont{Iino, Suzuki, and
  Ikushima}}]{Iino86}
\bibinfo{author}{\bibfnamefont{M.}~\bibnamefont{Iino}},
  \bibinfo{author}{\bibfnamefont{M.}~\bibnamefont{Suzuki}}, \bibnamefont{and}
  \bibinfo{author}{\bibfnamefont{A.~J.} \bibnamefont{Ikushima}},
  \bibinfo{journal}{J.\ Low Temp.\ Phys.} \textbf{\bibinfo{volume}{63}},
  \bibinfo{pages}{495} (\bibinfo{year}{1986}).

\bibitem[{\citenamefont{Andreev and Parshin}(1978)}]{AP}
\bibinfo{author}{\bibfnamefont{A.~F.} \bibnamefont{Andreev}} \bibnamefont{and}
  \bibinfo{author}{\bibfnamefont{A.~Y.} \bibnamefont{Parshin}},
  \bibinfo{journal}{JETP} \textbf{\bibinfo{volume}{48}}, \bibinfo{pages}{763}
  (\bibinfo{year}{1978}).

\bibitem[{\citenamefont{Keshishev et~al.}(1980)\citenamefont{Keshishev,
  Parshin, and Babkin}}]{Keshishev80}
\bibinfo{author}{\bibfnamefont{K.~O.} \bibnamefont{Keshishev}},
  \bibinfo{author}{\bibfnamefont{A.~Y.} \bibnamefont{Parshin}},
  \bibnamefont{and} \bibinfo{author}{\bibfnamefont{A.~V.}
  \bibnamefont{Babkin}}, \bibinfo{journal}{JETP Lett.}
  \textbf{\bibinfo{volume}{30}}, \bibinfo{pages}{56} (\bibinfo{year}{1980}).

\bibitem[{\citenamefont{Rolley et~al.}(1995)\citenamefont{Rolley, Guthmann,
  Shevalier, and Balibar}}]{Rolley95}
\bibinfo{author}{\bibfnamefont{E.}~\bibnamefont{Rolley}},
  \bibinfo{author}{\bibfnamefont{C.}~\bibnamefont{Guthmann}},
  \bibinfo{author}{\bibfnamefont{E.}~\bibnamefont{Shevalier}},
  \bibnamefont{and} \bibinfo{author}{\bibfnamefont{S.}~\bibnamefont{Balibar}},
  \bibinfo{journal}{J.\ Low Temp.\ Phys.} \textbf{\bibinfo{volume}{99}},
  \bibinfo{pages}{851} (\bibinfo{year}{1995}).

\bibitem[{\citenamefont{Bradley et~al.}(1995)\citenamefont{Bradley, Bunkov,
  Cousins, Enrico, Fisher, Follows, Gu\'{e}nault, Hayes, Pickett, and
  Sloan}}]{Bradley95}
\bibinfo{author}{\bibfnamefont{D.~I.} \bibnamefont{Bradley}},
  \bibinfo{author}{\bibfnamefont{Y.~M.} \bibnamefont{Bunkov}},
  \bibinfo{author}{\bibfnamefont{D.~J.} \bibnamefont{Cousins}},
  \bibinfo{author}{\bibfnamefont{M.~P.} \bibnamefont{Enrico}},
  \bibinfo{author}{\bibfnamefont{S.~N.} \bibnamefont{Fisher}},
  \bibinfo{author}{\bibfnamefont{M.~R.} \bibnamefont{Follows}},
  \bibinfo{author}{\bibfnamefont{A.~M.} \bibnamefont{Gu\'{e}nault}},
  \bibinfo{author}{\bibfnamefont{W.~M.} \bibnamefont{Hayes}},
  \bibinfo{author}{\bibfnamefont{G.~R.} \bibnamefont{Pickett}},
  \bibnamefont{and} \bibinfo{author}{\bibfnamefont{T.}~\bibnamefont{Sloan}},
  \bibinfo{journal}{Phys.\ Rev.\ Lett.} \textbf{\bibinfo{volume}{75}},
  \bibinfo{pages}{1887} (\bibinfo{year}{1995}).

\bibitem[{\citenamefont{Blaauwgeers et~al.}(2002)\citenamefont{Blaauwgeers,
  Eltsov, Eska, Finne, Haley, Krusius, Ruohio, Skrbek, and Volovik}}]{Eltsov02}
\bibinfo{author}{\bibfnamefont{R.}~\bibnamefont{Blaauwgeers}},
  \bibinfo{author}{\bibfnamefont{V.~B.} \bibnamefont{Eltsov}},
  \bibinfo{author}{\bibfnamefont{G.}~\bibnamefont{Eska}},
  \bibinfo{author}{\bibfnamefont{A.~P.} \bibnamefont{Finne}},
  \bibinfo{author}{\bibfnamefont{R.~P.} \bibnamefont{Haley}},
  \bibinfo{author}{\bibfnamefont{M.}~\bibnamefont{Krusius}},
  \bibinfo{author}{\bibfnamefont{J.~J.} \bibnamefont{Ruohio}},
  \bibinfo{author}{\bibfnamefont{L.}~\bibnamefont{Skrbek}}, \bibnamefont{and}
  \bibinfo{author}{\bibfnamefont{G.~E.} \bibnamefont{Volovik}},
  \bibinfo{journal}{Phys.\ Rev.\ Lett.} \textbf{\bibinfo{volume}{89}},
  \bibinfo{pages}{155301} (\bibinfo{year}{2002}).

\bibitem[{\citenamefont{Andreev}(1993)}]{Andreev93}
\bibinfo{author}{\bibfnamefont{A.~F.} \bibnamefont{Andreev}},
  \bibinfo{journal}{JETP Lett.} \textbf{\bibinfo{volume}{58}},
  \bibinfo{pages}{715} (\bibinfo{year}{1993}).

\bibitem[{\citenamefont{Leggett}(1975)}]{Leggett75}
\bibinfo{author}{\bibfnamefont{A.~J.} \bibnamefont{Leggett}},
  \bibinfo{journal}{Rev. Mod. Phys.} \textbf{\bibinfo{volume}{47}},
  \bibinfo{pages}{331} (\bibinfo{year}{1975}).

\bibitem[{\citenamefont{Bartkowiak et~al.}(1999)\citenamefont{Bartkowiak,
  Daley, Fisher, Gu\'{e}nault, Plenderleith, Haley, Pickett, and
  Skyba}}]{Bartkowiak99}
\bibinfo{author}{\bibfnamefont{M.}~\bibnamefont{Bartkowiak}},
  \bibinfo{author}{\bibfnamefont{S.~W.~J.} \bibnamefont{Daley}},
  \bibinfo{author}{\bibfnamefont{S.~N.} \bibnamefont{Fisher}},
  \bibinfo{author}{\bibfnamefont{A.~M.} \bibnamefont{Gu\'{e}nault}},
  \bibinfo{author}{\bibfnamefont{G.~N.} \bibnamefont{Plenderleith}},
  \bibinfo{author}{\bibfnamefont{R.~P.} \bibnamefont{Haley}},
  \bibinfo{author}{\bibfnamefont{G.~R.} \bibnamefont{Pickett}},
  \bibnamefont{and} \bibinfo{author}{\bibfnamefont{P.}~\bibnamefont{Skyba}},
  \bibinfo{journal}{Phys.\ Rev.\ Lett.} \textbf{\bibinfo{volume}{83}},
  \bibinfo{pages}{3462} (\bibinfo{year}{1999}).

\bibitem[{\citenamefont{Bartkowiak et~al.}(2002)\citenamefont{Bartkowiak,
  Fisher, Gu\'{e}nault, Pickett, Rogge, and Skyba}}]{Haley02}
\bibinfo{author}{\bibfnamefont{M.}~\bibnamefont{Bartkowiak}},
  \bibinfo{author}{\bibfnamefont{S.~N.} \bibnamefont{Fisher}},
  \bibinfo{author}{\bibfnamefont{A.~M.} \bibnamefont{Gu\'{e}nault}},
  \bibinfo{author}{\bibfnamefont{G.~R.} \bibnamefont{Pickett}},
  \bibinfo{author}{\bibfnamefont{M.~C.} \bibnamefont{Rogge}}, \bibnamefont{and}
  \bibinfo{author}{\bibfnamefont{P.}~\bibnamefont{Skyba}},
  \bibinfo{journal}{J.\ Low Temp.\ Phys.} \textbf{\bibinfo{volume}{126}},
  \bibinfo{pages}{533} (\bibinfo{year}{2002}).

\bibitem[{\citenamefont{Bartkowiak et~al.}(2000)\citenamefont{Bartkowiak,
  Fisher, Gu\'{e}nault, Haley, Plenderleith, Pickett, and
  Skyba}}]{Bartkowiak00}
\bibinfo{author}{\bibfnamefont{M.}~\bibnamefont{Bartkowiak}},
  \bibinfo{author}{\bibfnamefont{S.~N.} \bibnamefont{Fisher}},
  \bibinfo{author}{\bibfnamefont{A.~M.} \bibnamefont{Gu\'{e}nault}},
  \bibinfo{author}{\bibfnamefont{R.~P.} \bibnamefont{Haley}},
  \bibinfo{author}{\bibfnamefont{G.~N.} \bibnamefont{Plenderleith}},
  \bibinfo{author}{\bibfnamefont{G.~R.} \bibnamefont{Pickett}},
  \bibnamefont{and} \bibinfo{author}{\bibfnamefont{M.}~\bibnamefont{Skyba}},
  \bibinfo{journal}{Physica B} \textbf{\bibinfo{volume}{284}},
  \bibinfo{pages}{240} (\bibinfo{year}{2000}).

\bibitem[{\citenamefont{Wheatley}(1975)}]{Wheatley75}
\bibinfo{author}{\bibfnamefont{J.~C.} \bibnamefont{Wheatley}},
  \bibinfo{journal}{Rev.\ Mod.\ Phys.} \textbf{\bibinfo{volume}{47}},
  \bibinfo{pages}{415} (\bibinfo{year}{1975}).

\bibitem[{\citenamefont{Osheroff and Brinkmann}(1974)}]{OsheroffBrinkmann74}
\bibinfo{author}{\bibfnamefont{D.~D.} \bibnamefont{Osheroff}} \bibnamefont{and}
  \bibinfo{author}{\bibfnamefont{W.~F.} \bibnamefont{Brinkmann}},
  \bibinfo{journal}{Phys.\ Rev.\ Lett.} \textbf{\bibinfo{volume}{32}},
  \bibinfo{pages}{584} (\bibinfo{year}{1974}).

\bibitem[{\citenamefont{Webb et~al.}(1974)\citenamefont{Webb, Kleinberg, and
  Wheatley}}]{Webb74}
\bibinfo{author}{\bibfnamefont{R.~A.} \bibnamefont{Webb}},
  \bibinfo{author}{\bibfnamefont{R.~L.} \bibnamefont{Kleinberg}},
  \bibnamefont{and} \bibinfo{author}{\bibfnamefont{J.~C.}
  \bibnamefont{Wheatley}}, \bibinfo{journal}{Phys.\ Rev.\ Lett.}
  \textbf{\bibinfo{volume}{33}}, \bibinfo{pages}{145} (\bibinfo{year}{1974}).

\bibitem[{\citenamefont{Bozler et~al.}(1974)\citenamefont{Bozler, Bernier,
  Gully, Richardson, and Lee}}]{Bozler74}
\bibinfo{author}{\bibfnamefont{H.~M.} \bibnamefont{Bozler}},
  \bibinfo{author}{\bibfnamefont{M.~E.~R.} \bibnamefont{Bernier}},
  \bibinfo{author}{\bibfnamefont{W.~J.} \bibnamefont{Gully}},
  \bibinfo{author}{\bibfnamefont{R.~C.} \bibnamefont{Richardson}},
  \bibnamefont{and} \bibinfo{author}{\bibfnamefont{D.~M.} \bibnamefont{Lee}},
  \bibinfo{journal}{Phys.\ Rev.\ Lett.} \textbf{\bibinfo{volume}{32}},
  \bibinfo{pages}{875} (\bibinfo{year}{1974}).

\bibitem[{\citenamefont{Osheroff}(1974)}]{Osheroff74}
\bibinfo{author}{\bibfnamefont{D.~D.} \bibnamefont{Osheroff}},
  \bibinfo{journal}{Phys.\ Rev.\ Lett.} \textbf{\bibinfo{volume}{33}},
  \bibinfo{pages}{1009} (\bibinfo{year}{1974}).

\bibitem[{\citenamefont{Leggett}(1972)}]{Leggett72}
\bibinfo{author}{\bibfnamefont{A.~J.} \bibnamefont{Leggett}},
  \bibinfo{journal}{Phys.\ Rev.\ Lett.} \textbf{\bibinfo{volume}{29}},
  \bibinfo{pages}{1227} (\bibinfo{year}{1972}).

\bibitem[{\citenamefont{Halperin et~al.}(1976)\citenamefont{Halperin, Archie,
  Rasmussen, Alvesalo, and Richardson}}]{Halperin76}
\bibinfo{author}{\bibfnamefont{W.~P.} \bibnamefont{Halperin}},
  \bibinfo{author}{\bibfnamefont{C.~N.} \bibnamefont{Archie}},
  \bibinfo{author}{\bibfnamefont{F.~B.} \bibnamefont{Rasmussen}},
  \bibinfo{author}{\bibfnamefont{T.~A.} \bibnamefont{Alvesalo}},
  \bibnamefont{and} \bibinfo{author}{\bibfnamefont{R.~C.}
  \bibnamefont{Richardson}}, \bibinfo{journal}{Phys.\ Rev.\ B}
  \textbf{\bibinfo{volume}{13}}, \bibinfo{pages}{2124} (\bibinfo{year}{1976}).

\bibitem[{\citenamefont{Alvesalo et~al.}(1980)\citenamefont{Alvesalo,
  Haavasoja, Manninen, and Soinne}}]{Alvesalo80}
\bibinfo{author}{\bibfnamefont{T.~A.} \bibnamefont{Alvesalo}},
  \bibinfo{author}{\bibfnamefont{T.}~\bibnamefont{Haavasoja}},
  \bibinfo{author}{\bibfnamefont{M.~T.} \bibnamefont{Manninen}},
  \bibnamefont{and} \bibinfo{author}{\bibfnamefont{A.~T.}
  \bibnamefont{Soinne}}, \bibinfo{journal}{Phys.\ Rev.\ Lett.}
  \textbf{\bibinfo{volume}{44}}, \bibinfo{pages}{1076} (\bibinfo{year}{1980}).

\bibitem[{\citenamefont{Greywall}(1986)}]{Greywall86}
\bibinfo{author}{\bibfnamefont{D.~S.} \bibnamefont{Greywall}},
  \bibinfo{journal}{Phys.\ Rev.\ B} \textbf{\bibinfo{volume}{33}},
  \bibinfo{pages}{7520} (\bibinfo{year}{1986}).

\bibitem[{\citenamefont{Yip and Leggett}(1986)}]{Leggett86}
\bibinfo{author}{\bibfnamefont{S.}~\bibnamefont{Yip}} \bibnamefont{and}
  \bibinfo{author}{\bibfnamefont{A.~J.} \bibnamefont{Leggett}},
  \bibinfo{journal}{Phys.\ Rev.\ Lett.} \textbf{\bibinfo{volume}{57}},
  \bibinfo{pages}{345} (\bibinfo{year}{1986}).

\bibitem[{\citenamefont{Leggett and Yip}(1990)}]{Leggett90}
\bibinfo{author}{\bibfnamefont{A.~J.} \bibnamefont{Leggett}} \bibnamefont{and}
  \bibinfo{author}{\bibfnamefont{S.~K.} \bibnamefont{Yip}},
  \emph{\bibinfo{title}{in Helium Three, edited by W.P.\,Halperin and
  L.P.\,Pitaevskii}} (\bibinfo{publisher}{North Holland},
  \bibinfo{address}{Amsterdam}, \bibinfo{year}{1990}).

\bibitem[{\citenamefont{Grilly}(1971)}]{Grilly}
\bibinfo{author}{\bibfnamefont{E.~R.} \bibnamefont{Grilly}},
  \bibinfo{journal}{J.\ Low Temp.\ Phys.} \textbf{\bibinfo{volume}{4}},
  \bibinfo{pages}{615} (\bibinfo{year}{1971}).

\bibitem[{\citenamefont{Kopnin}(1987)}]{Kopnin87}
\bibinfo{author}{\bibfnamefont{N.~B.} \bibnamefont{Kopnin}},
  \bibinfo{journal}{Sov.\ Phys.\ JETP} \textbf{\bibinfo{volume}{65}},
  \bibinfo{pages}{1187} (\bibinfo{year}{1987}).

\bibitem[{\citenamefont{Buchanan et~al.}(1986)\citenamefont{Buchanan, Swift,
  and Wheatley}}]{Buchanan86}
\bibinfo{author}{\bibfnamefont{D.~S.} \bibnamefont{Buchanan}},
  \bibinfo{author}{\bibfnamefont{G.~W.} \bibnamefont{Swift}}, \bibnamefont{and}
  \bibinfo{author}{\bibfnamefont{J.~C.} \bibnamefont{Wheatley}},
  \bibinfo{journal}{Phys.\ Rev.\ Lett.} \textbf{\bibinfo{volume}{57}},
  \bibinfo{pages}{341} (\bibinfo{year}{1986}).

\bibitem[{\citenamefont{Stern}(2010)}]{Stern10}
\bibinfo{author}{\bibfnamefont{A.}~\bibnamefont{Stern}},
  \bibinfo{journal}{Nature} \textbf{\bibinfo{volume}{464}},
  \bibinfo{pages}{187} (\bibinfo{year}{2010}).

\end{thebibliography}

\newpage
\end{document}